\documentclass[pra,floatfix,twocolumn,showpacs,superscriptaddress]{revtex4-1}

\usepackage{amsmath}
\usepackage{amsfonts}
\usepackage[linktocpage,colorlinks=true,linkcolor=blue,citecolor=blue]{hyperref}
\usepackage{graphicx}
\usepackage{float}
\usepackage[caption=false]{subfig}
\begin{document}
{
\title{Relative Volume of Separable Bipartite States}
\author{Rajeev Singh}
\affiliation{Optics \& Quantum Information Group, 
The Institute of Mathematical Sciences, CIT Campus, Taramani,
Chennai 600113, India.}
\affiliation{Max-Planck-Institut f\"{u}r Physik komplexer Systeme,
01187 Dresden, Germany.}
\author{Ravi Kunjwal}
\affiliation{Optics \& Quantum Information Group, 
The Institute of Mathematical Sciences, CIT Campus, Taramani,
Chennai 600113, India.}
\author{R. Simon}
\affiliation{Optics \& Quantum Information Group, 
The Institute of Mathematical Sciences, CIT Campus, Taramani,
Chennai 600113, India.}
\date{\today}
\begin{abstract}
Every choice of an orthonormal frame in the $d$-dimensional Hilbert space of a
system corresponds to one set of all mutually commuting density matrices
or, equivalently, a classical statistical state space of the system; the quantum state
space itself can thus be profitably viewed as an $SU(d)$ orbit of classical state
spaces, one for each orthonormal frame. We exploit this connection to
study the relative volume of separable states of a bipartite quantum
system. While the two-qubit case is studied in considerable analytic detail, for
higher dimensional systems we fall back on Monte Carlo.
Several new insights seem to emerge from our study.
\end{abstract}
\maketitle
\section{Introduction}
States of a quantum system are represented by density
operators (positive semidefinite unit-trace operators acting
on a $d$-dimensional Hilbert space ${\cal H}_d$).
The set of all density operators of such a $d$-dimensional ($d$-level) system constitutes a convex subset of
$\mathbb{R}^{d^{\, 2}-1}$; this is
the {\em state space} ({\em generalized Bloch sphere}) $\Omega_d$ of 
the quantum system.
An understanding of the geometry of the state space
is of fundamental
importance~\cite{Bengtsson2006,Bengtsson2013}.
The state space of a two-level system or qubit is the well-known Bloch
(or Poincar\'{e}) sphere, while the generalized Bloch sphere of
higher dimensional
system is much richer, and more complex to visualize and analyze.
Recently a limited analysis of the cross-sections of the state
space of the three-level system (qutrit) has been
performed~\cite{Goyal2011,Sarbicki2013}.

When $d$ is non-prime, it is possible that the system is
composite, i.e. made up of two or more subsystems.
For example, a $4$-dimensional system could be a single quantum system
with four levels or a pair of two-level systems or qubits.
In the latter case of composite system, the issue of separability becomes
important,
entanglement being a characteristic feature of quantum theory of
composite systems, and
a key resource in quantum information processing~\cite{Horodecki2009}.
An understanding of the separability property of states is therefore important from both
foundational and application perspectives.
We would like to understand the geometry of separable states and know
how much of the state space is entangled.

The issue regarding the relative volume of the set of all separable
states was considered in the seminal work of \.{Z}yczkowski
{\em et. al.}~\cite{Zyczkowski1998}.
It was not only shown that the set of separable states has
non-zero volume, but also analytical lower and upper bounds were obtained
for the two-qubit and the qubit-qutrit cases.
They also argued that all states in a sufficiently small neighborhood
of the maximally mixed state are separable, and conjectured that the
volume of the separable region decreases exponentially with Hilbert space
dimension. Different aspects of this issue have been addressed by
other authors~\cite{Zyczkowski1999,Slater1999,Kendon2002a}.
Vidal and Tarrach~\cite{Vidal1999} generalized the result to obtain
an analytical lower bound on this volume for multi-partite systems,
showing that it is non-zero.
Verstraete {\em et. al.}~\cite{Verstraete2001} gave an improved lower
bound 
on the volume of the separable region for the two-qubit system.
More recently, significant contribution has been made to the
understanding of the generalized two-qubit Hilbert-Schmidt separability
probabilities by Slater\,\cite{Slater2012,Slater2013}.
In similar work on pure states it was shown that typical or generic pure states of
multiple-qubit systems are highly entangled, while having low amounts
of pairwise entanglement~\cite{Kendon2002}.
Regarding the issue of geometry of state space, the geometry of
Bell-diagonal states for two-qubit systems in the context of quantum
discord has been addressed recently~\cite{Lang2010}.

The genesis of the present work is the following. During a recent reading of the seminal work of \.{Z}yczkowski {\em et. al.}~\cite{Zyczkowski1998}---a paper we had indeed read more than once earlier---the
following observation by these authors somehow captured our attention:``{\em Our numerical results agree with these bounds, but to our surprise the probability that a mixed state $\rho \in \mathcal{H}_2 \otimes \mathcal{H}_2$
is separable exceeds 50\%}.'' Their paper established an interesting analytical lower bound of $0.302$ for the probability of separability (fractional volume of separable states) of a two-qubit system
(and an analytical upper bound of $0.863$), but on numerical (Monte Carlo) estimation they found it to actually exceed $50\%$ and assume $0.632$. We could not resist asking ourselves the following question:
{\em Could there be a ground to `anticipate' this value in excess of 50\%?} It is this question that marked the humble beginning of the present work.

The quantum (statistical) state space of a two-state system or qubit is simply 
the Bloch (Poincar\'{e}) sphere---a unit ball $B_3 \subset \mathbb{R}^3$ 
centered at the origin; but for $d\geq 3$ the generalized Bloch `sphere' has a 
much richer structure. It is a convex body $\Omega_d \subset \mathbb{R}^{d^{\, 
2}-1}$ determined by $CP^{{\,d}-1}$ worth of pure states as extremals, this 
$2(d-1)$-parameter family of pure states being `sprinkled over' the 
$(d^{\,2}-2)$-dimensional boundary of $\Omega_d$. In contrast, the classical 
(statistical) state space of a $d$-state system is extremely simple, for all 
$d$. Indeed, it is the regular simplex $\Delta_{{\,d}-1} \subset 
\mathbb{R}^{{\,d}-1}$, the convex body defined by $d$ equidistant vertices or 
extremals (the classical pure states). {\em The quantum state space itself can 
be profitably viewed, for all $d$, as the union of the $SU(d)$ orbit ~$\Gamma$ 
of simplices (classical state spaces) $\Delta_{{\,d}-1}$}. This fact is 
fundamental to both our point of view and analysis in this work. Every set of 
all mutually commuting $d \times d$ density matrices constitutes one classical 
state space or simplex $\Delta_{{\,d}-1}$, a point in the orbit, and choice of 
a set or frame of orthonormal unit vectors (more properly, unit rays) labels 
different points on the orbit. Thus, the orbit $\Gamma$ is exactly as large as 
the coset space $U(d)/[U(1)\times U(1) \times \cdots \times U(1)]$, a 
particular case of (complex) Stiefel manifold. The volume of 
 $\Omega_d$ is 
thus the product of the volume of the simplex 
$\Delta_{{\,d}-1}$ and the volume of the 
$(d^2-d)$-dimensional orbit $U(d)/[U(1)\times U(1) \times \cdots \times 
U(1)]$, the later volume being determined by the measure inherited from the 
Haar (uniform) measure on the unitary group $SU(d)$.

Our interest here is the $d_1 \times d_2$ bipartite system, and therefore the 
relevant simplex is $\Delta_{{\,d}_1 d_2 - 1}$ and  the dimension of 
the orbit $\Gamma$ of orthonormal frames is $d_1d_2(d_1d_2-1)$. {\em 
Separability issues are invariant under local unitaries} $U_{d_1}\otimes 
U_{d_2}$, so it is sufficient to restrict attention to the local unitarily 
inequivalent frames. This removes $d_1^{\,2}+d_2^{\,2}-2$ parameters, and so 
we are left with $(d_1^{\,2}-1)(d_2^{\,2}-1)-d_1d_2+1$ parameters needed to 
label points on the orbit of local-unitarily inequivalent orthonormal frames 
or simplices $\Delta_{{\,d}_1d_2-1}$, this number evaluating to $6$ for the 
two-qubit systems and to $56$ for the two-qutrit systems.

We now have all the ingredients to describe our approach to the problem of 
(fractional) volume of separable states in more precise terms. Considering the 
full state space $\Omega_{d_1d_2}$ of the $d_1\times d_2$ bipartite system as an orbit $\Gamma$ 
of simplices $\Delta_{{\,d}_1d_2-1}$, let $\xi$ denote the collection of 
variables, say $k$ in number, needed to label points on the orbit or manifold 
$\Gamma$, i.e., for each $\xi \in \Gamma$ we have an orthonormal basis of 
$d_1d_2$-dimensional vectors  and an associated simplex 
$\Delta_{{\,d}_1d_2-1}(\xi)$ of mutually commuting density operators. The 
volume of simplex $\Delta_{{\,d}_1d_2-1}(\xi)$ is independent of $\xi$; this 
fact is trivially obvious, but proves important for our present purpose. For 
each $\xi$, a convex subset of $\Delta_{{\,d}_1d_2-1}(\xi)$, whose volume is 
not independent of $\xi$, is separable. Let $f(\xi)$ represent the fractional 
$(d_1d_2-1)$-dimensional volume of this convex subset of 
$\Delta_{{\,d}_1d_2-1}(\xi)$. The uniform Haar measure on the unitary group 
$SU(d_1d_2)$ induces a measure or probability $p(\xi)$ on the orbit $\Gamma$. 
Clearly, the fractional volume of separable states for the full space is given 
by
\begin{equation}\label{vsep}
 V^{\rm sep}/V^{\rm tot}\equiv v^{\rm sep}=\int{d^k\xi~p(\xi)~f(\xi)}.
\end{equation}
An immediate and important implication of this rendering of relative 
volume of separable states is this:
 {\em Should it turn out that $f(\xi)\geq a>0, ~\forall \xi \in \Gamma$,
then $v^{\rm sep}$ is trivially bounded from below by $a$}. For
the two-qubit system we shall indeed show that $a=0.5$, thus 
reconciling the `surprise' 
element which acted as the `seed` for the present work, as noted earlier.

Two remarks are in order in respect of our analysis leading to
Eq.~(\ref{vsep}), one in respect of choice of measure over the simplex
and the other regarding the fact that the simplices corresponding to
two distinct points of the orbit $\Gamma$ are not necessarily
disjoint.

\noindent
{\bf Remark 1}: There exists a natural volume measure for $\Gamma$
arising from the very fact that it is an $SU(d_1 d_2)$ orbit.
But the situation in respect of the simplex $\Delta_{{\,d}_1d_2-1}$ is quite
different.
There seems to exist no fundamental mathematical principle to pick one
unique or distinguished measure on $\Delta_{{\,d}_1d_2-1}$, and therefore the
choice seems to be ultimately a matter of taste or point of view.
However, the action of the permutation group $S_{{\,d}_1 d_2}$ on the simplex 
 $\Delta_{{\,d}_1d_2-1}$, through permutation of its vertices,  
renders $\Delta_{{\,d}_1d_2-1}$
the union of $(d_1 d_2)!$ mutually equivalent fundamental domains.
Thus the complete freedom in choice of measure applies to one
fundamental domain, of fractional volume $1 / (d_1 d_2)!$. The measure is 
transferred to the other copies of the fundamental domain by the natural action of the permutation group $S_{d_1d_2}$ . 
The choice of \.{Z}yczkowski {\em et. al.} is the uniform measure, the
one inherited by embedding $\Delta_{{\,d}_1d_2-1}$ in the Euclidean space 
$\mathbb{R}^{d_1 d_2 - 1}$.
Other measures have been motivated and used
in~\cite{Zyczkowski1999,Slater1999}.
Since the present work was inspired by that  of \.{Z}yczkowski
{\em et. al.}, we stick to their measure.

\noindent
{\bf Remark 2}: The different simplices on the orbit $\Gamma$ are not
necessarily disjoint.
As is readily seen, the intersection is however restricted to those
points of the simplex which correspond to density matrices with
degenerate spectrum.
For instance, in the case of a qutrit such points correspond precisely
to the bisectors of the equilateral triangle, the $2$-simplex
$\Delta_2$.
Since such points of zero measure contribute neither to the total
volume of the simplex nor to that of its separable convex subset, the
fact that the simplices are not disjoint affects in no way the
development leading to Eq.~(\ref{vsep}).

The content of the paper is organized as follows. In Section II we present 
details of the two-qubit system, and this is
 followed by numerical Monte Carlo analysis for 
higher dimensional systems in Section III. In Section IV we make a few 
observations on the separable volume for the qubit-qutrit system. Section V 
concludes with a comment on the volume and `effective radius' of the separable 
region for higher dimensional quantum systems. The final Section VI summarizes 
our results. 

\section{Two Qubit System} 

The state space of a two-qubit system 
$\Omega_4$ corresponds to positive semidefinite unit-trace operators on the 
$4$-dimensional Hilbert space, and in the present scheme can be symbolically 
expressed as $\Omega_4 \sim \Gamma_{22} \times \Delta_{\, 3}$. But this 
$15$-parameter convex set $\Omega_4$ should not be viewed as the Cartesian 
product of the two sets $\Gamma_{22}$ and $\Delta_{\, 3}$, but rather as the 
union of $3$-simplices (tetrahedra) $\Delta_{\, 3}$ parameterized by the 
$12$-parameter manifold of frames $\Gamma_{22} = U(4)/[U(1) \times U(1) \times 
U(1) \times U(1)]$. We first describe the 3-simplex $\Delta_{\, 3}$ comprising 
probabilities $\{p_j\}$, $\sum_{j=1}^4 p_j = 1, ~p_j \ge 0, ~j = 1,\cdots,4$. 
Since $\Delta_{\, 3}$ resides in a $3$-dimensional Cartesian real space, it is 
both desirable and instructive to pictorially visualize this simplex along 
with its convex subset of separable states. We can explicitly picture the 
separable set corresponding to any selected frame using the following change 
of variables from the four $p_j$'s {\em constrained} by $\sum p_j = 1$ to 
three {\em independent} Cartesian variables $x,y,z$: 
\begin{equation} 
\begin{aligned} p_1 &= (1+x+y+z) / 4, \\ p_2 &= (1+x-y-z) / 4,\\ p_3 &= 
(1-x+y-z) / 4, \\ p_4 &= (1-x-y+z) / 4. \end{aligned} \end{equation} 
The 
situation where one particular $p_j = 0,~j = 1, \cdots, 4$ is seen to 
correspond to one of the four faces of the tetrahedron or 3-simplex 
$\Delta_{\, 3}$ in the three-dimensional $xyz$ space with vertices at 
$(1,1,1)$, $(1,-1,-1)$, $(-1,1,-1)$ and $(-1,-1,1)$ (see 
Fig.~\ref{SeparableSet}). The six edges correspond to pairs of $p_j$'s 
vanishing, and the vertices to only one nonvanishing $p_j$. In this way we 
associate a tetrahedron with {\em every} set of all mutually commuting density 
matrices determined by choice of a frame of four orthonormal pure states $\{ 
|\Psi_k \rangle \}$, $\langle \Psi_j | \Psi_k \rangle = \delta_{jk}$.

\subsection{A Special Two-parameter Family of Frames}
\begin{figure*}
  \centering
  \subfloat[$(0,0)$]{\includegraphics[clip, width=0.33\linewidth]{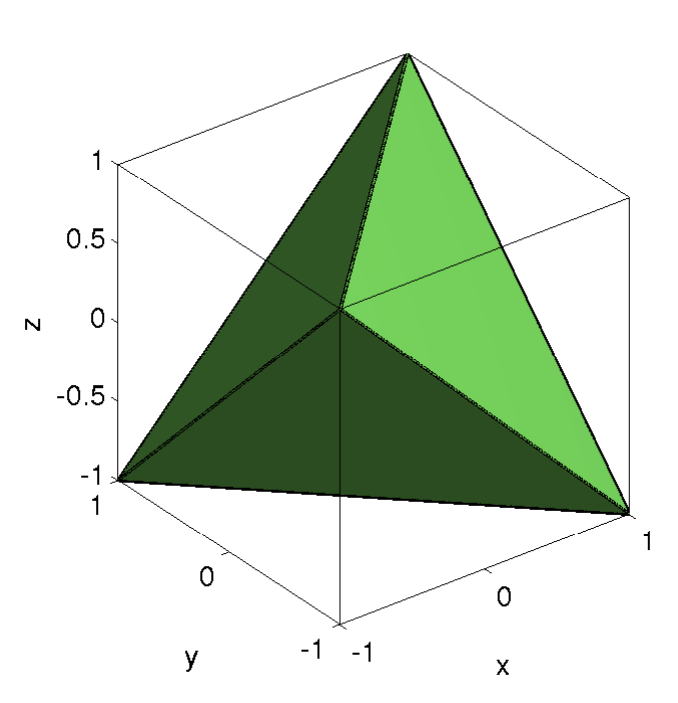}}
  ~
  \subfloat[$(0,\pi/8)$]{\includegraphics[clip,width=0.33\linewidth]{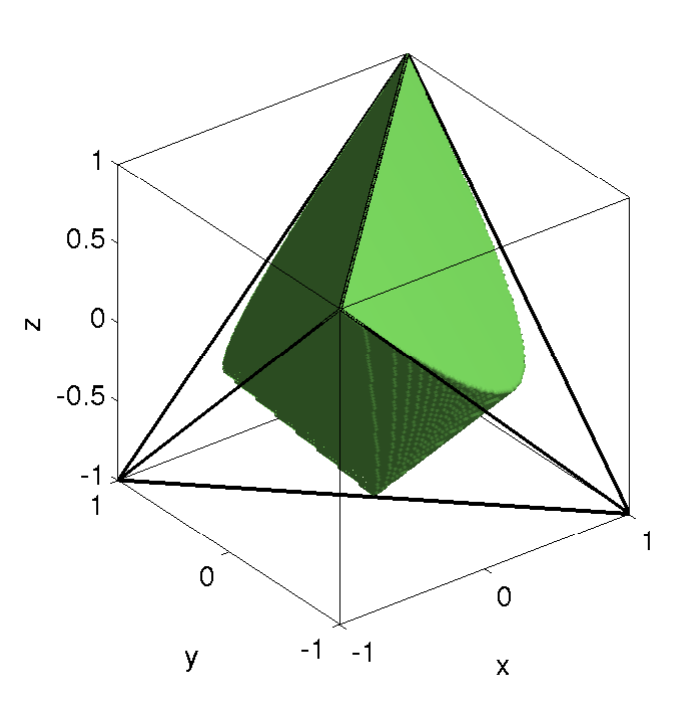}}
  ~
  \subfloat[$(0,\pi/4)$]{\includegraphics[clip,width=0.33\linewidth]{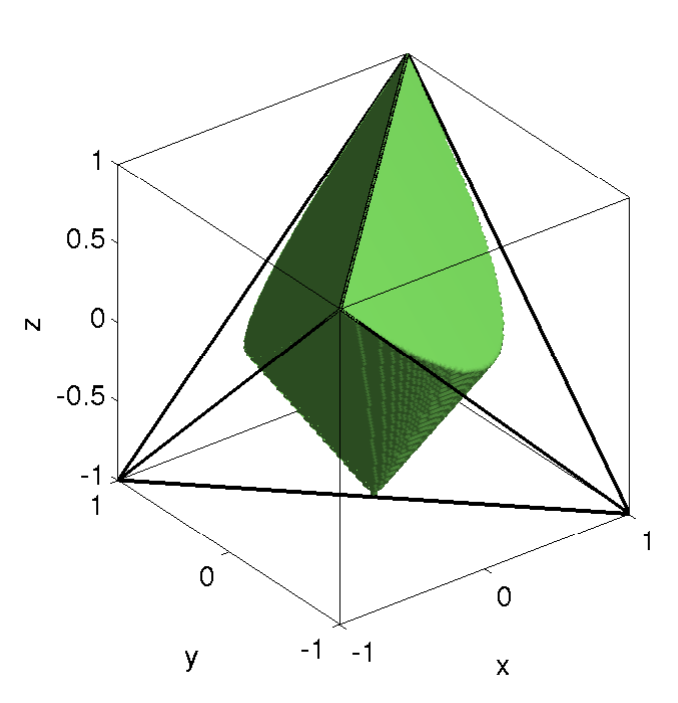}}
  \\
  \subfloat[$(\pi/8,\pi/8)$]{\includegraphics[clip,width=0.33\linewidth]{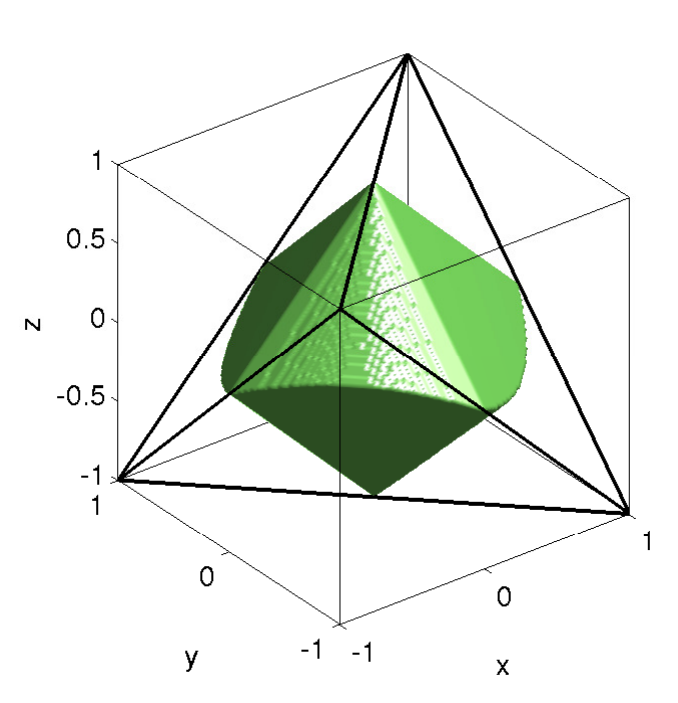}}
  ~
  \subfloat[$(\pi/8,\pi/4)$]{\includegraphics[clip,width=0.33\linewidth]{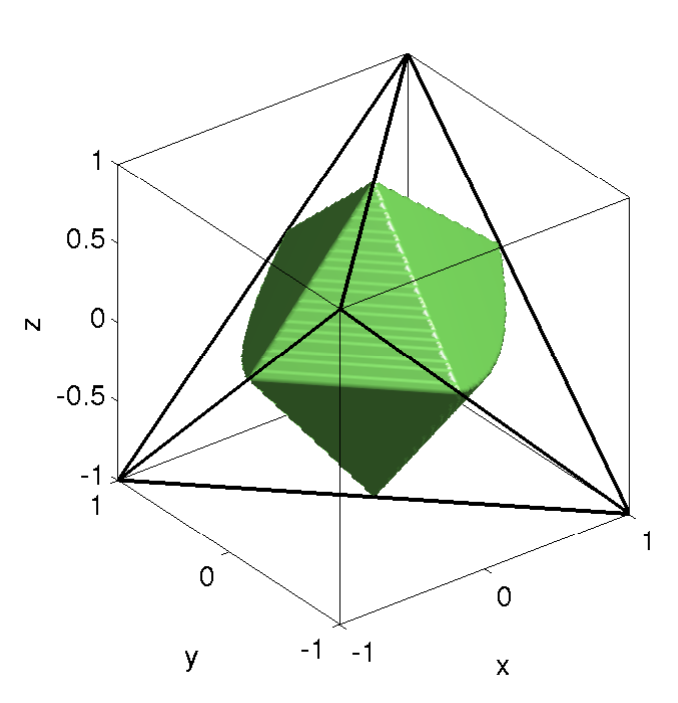}}
  ~
  \subfloat[$(\pi/4,\pi/4)$]{\includegraphics[clip,width=0.33\linewidth]{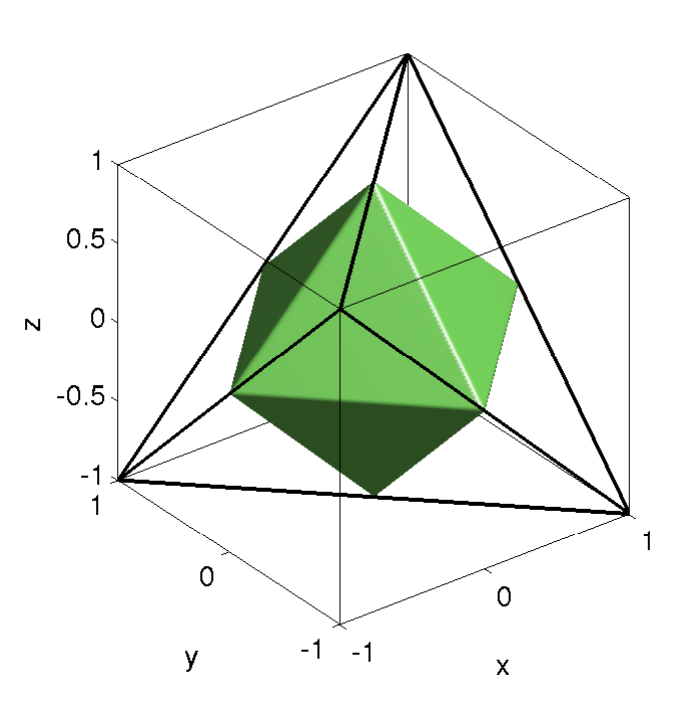}}
  \caption{(Color online) Separable regions for different values of $(\theta,\alpha)$.
  Each tetrahedron represents a set of all density matrices with common 
  eigenvectors.
  The volume enclosed by the shaded surface shows the separable
  region for the given frame.
  We find that the separable set is the entire tetrahedron for
  $(\theta,\alpha) = (0,0)$ and is an octahedron for
  $(\theta,\alpha) = (\pi/4,\pi/4)$ as expected.
  For other values of $(\theta,\alpha)$ we find the separable set to
  be the tetrahedron limited by planes and conic surfaces.
  }
  \label{SeparableSet}
\end{figure*}

\begin{figure}
  \centering
  \includegraphics[clip, width=1.0\linewidth]{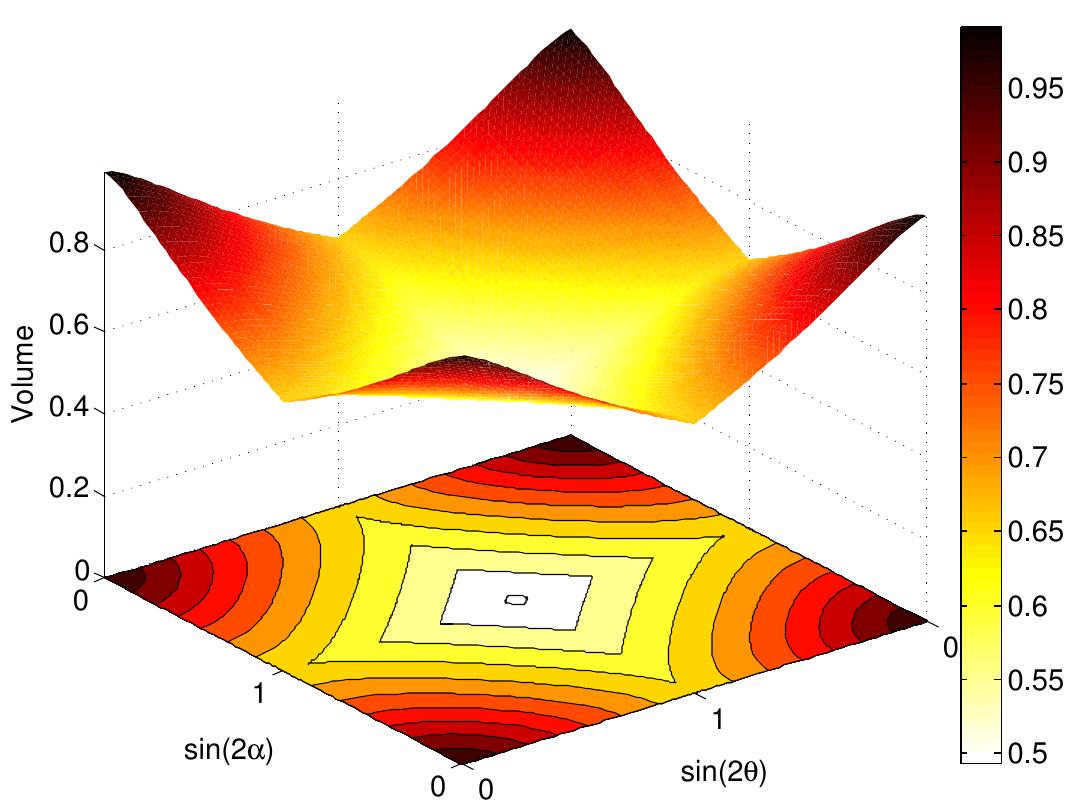}
  \caption{(Color online) Volume of separable states as a function of the two
  entanglement parameters $\sin 2\theta, \sin 2\alpha$ in the case of
  the two-parameter family of frames.
   It is seen  that the volume is minimum for $(\pi/4,\pi/4)$ which
  corresponds to the Bell or magic  frame.
  }
  \label{SeparableVolume}
\end{figure}

Before we discuss the general parameterization of the 12-parameter manifold
$\Gamma_{22}$ of two-qubit frames, for clarity of
presentation we consider first a special two-parameter
family of locally inequivalent frames which
are obtained as two orthonormal linear combinations
 within the computational basis pair $\{|00\rangle, |11\rangle\}$ and two within the
pair $\{|01\rangle, |10\rangle\}$\,:
\begin{equation}
\begin{aligned}
& | \Psi_1 \rangle = \cos \theta\ |00 \rangle + \sin \theta\ |11 \rangle, \\
& | \Psi_2 \rangle = \sin \theta\ |00 \rangle - \cos \theta\ |11 \rangle, \\
& | \Psi_3 \rangle = \cos \alpha\ |01 \rangle + \sin \alpha\ |10 \rangle, \\
& | \Psi_4 \rangle = \sin \alpha\ |01 \rangle - \cos \alpha\ |10 \rangle.
\end{aligned}
\end{equation}
These frames can be viewed, in an obvious manner, as a two-parameter
generalization of the Bell or magic frame of maximally entangled states.
Indeed, the Bell basis corresponds to $\theta = \pi/4 = \alpha$.
The entanglement of the first two states is determined by $\sin(2 \theta)$
while that of the next two by $\sin(2 \alpha)$.
That there are only two parameters is an immediate consequence of our
forbidding superposition across the two pairs of vectors,
$\{|00\rangle, |11\rangle\}$ and $\{|01\rangle, |10\rangle\}$.
It is readily verified that if one constructs any orthonormal pair of vectors
as linear combinations of $|00\rangle$ and $|11\rangle$, both would
have one and the same measure of entanglement; 
the same is true of $|01\rangle$ and $|10\rangle$ as well.
For this special parameterization the density matrix corresponding to
a given point $\{p_j\}$ in $\Delta_3$ is
\begin{widetext}
\begin{equation}
\begin{aligned}
\rho (\{p_i\}) &= \sum_{j=1}^4 p_j \ |\Psi_j \rangle \langle \Psi_j| \\
&=
\begin{pmatrix}
    p_1 \cos^2 \theta + p_2 \sin^2 \theta & 0 &
        0 & (p_1-p_2) \sin \theta \cos \theta \\
    0 & p_3 \cos^2 \alpha + p_4 \sin^2 \alpha &
        (p_3-p_4) \sin \alpha \cos \alpha & 0 \\
    0 & (p_3-p_4) \sin \alpha \cos \alpha &
        p_3 \sin^2 \alpha + p_4 \cos^2 \alpha & 0 \\
    (p_1-p_2) \sin \theta \cos \theta & 0 &
        0 & p_1 \sin^2 \theta + p_2 \cos^2 \theta
\end{pmatrix}.
\end{aligned}
\end{equation}
\end{widetext}
It is well known that positivity under partial transpose (PPT) is both
a necessary and sufficient condition for separability of the
qubit-qubit system~\cite{Peres1996,Horodecki1996}.
Since the partial transpose of the above matrix is a direct sum of $2\times 2$ matrices, the
condition for separability attains a simple (quadratic) form in $\{p_j\}$ (or $x,y,z$):
\begin{equation}
\begin{aligned}
(p_1^2 + p_2^2) \sin^2 \theta \cos^2 \theta
&+ p_1 p_2 (\sin^4 \theta + \cos^4 \theta) \\
&- (p_3 - p_4)^2 \sin^2 \alpha \cos^2 \alpha \ge 0, \\
(p_3^2 + p_4^2) \sin^2 \alpha \cos^2 \alpha
&+ p_3 p_4 (\sin^4 \alpha + \cos^4 \alpha) \\
&- (p_1 - p_2)^2 \sin^2 \theta \cos^2 \theta \ge 0.
\end{aligned}
\end{equation}
It is clear that (saturation of) these separability inequalities, for a given
numerical pair $(\theta, \alpha)$,
corresponds to surfaces that are quadratic in the $xyz$ space.
For special values of the parameters one or both of these quadratic
surfaces might factorize to give planes.
Thus, the boundaries of the separable region of $\Delta_3$, for any choice of $(\theta,\alpha)$, consist entirely of
quadratic and planar surfaces.

In Fig.~\ref{SeparableSet} we picture the separable region (inside the
tetrahedron) for a few selected values of $(\theta, \alpha)$.
The Bell or magic frame which corresponds to
$\theta = \alpha = \pi/4$ is shown as the last and sixth (as is well known, the separable region
is an octahedron in this case).
We numerically estimate the volume of the separable region for each
value of $(\theta, \alpha)$, and the result is pictured in
Fig.~\ref{SeparableVolume} in the $(\sin 2\theta, \sin 2\alpha)$ plane.
Clearly, the volume decreases with increasing `entanglement of the frame'.
Since the volume of the octahedron is exactly half the volume of the
tetrahedron of which it is a convex subset, the ratio of the volume of
separable states to the total volume $V^{{\rm sep}} / V^{{\rm tot}} = 0.5$ for the
Bell frame.
For every other frame in this two-parameter family this ratio is
larger, as is evident from Fig.~\ref{SeparableVolume}.

\subsection{Parameterization of $\Gamma_{22}$}
Having looked at a special two-parameter family of frames in some
detail, now we move on to parameterization of the full orbit $\Gamma_{22}$
of two-qubit frames, {\em modulo local unitaries}.
To this end, we expand a generic set of orthonormal two-qubit
vectors $\{ |\Psi_k \rangle \}$ in the computational basis:
\begin{equation}
|\Psi_k \rangle = \sum_{a,b=1}^2 C_{\, ab}^{(k)} \ |a \rangle_A
    \otimes |b \rangle_B, \qquad k=1,2,3,4.
\end{equation}
Orthonormality of the set $\{ |\Psi_k \rangle \}$ reads as
the trace-orthonormality condition
\begin{equation}
\langle \Psi_j | \Psi_k \rangle = {\rm Tr}( C^{(j)\dagger} C^{(k)} )
    = \delta_{jk}
 \label{TraceOrthonormality}
\end{equation}
on the corresponding set of $2\times2$ matrices $\{ C^{(k)} \}$ of expansion
coefficients.
Clearly, {\em quadruples of complex $2 \times 2$ matrices
$\{ C^{(k)} \}$ meeting the requirement~(\ref{TraceOrthonormality})
are in one-to-one correspondence with ONB's or frames in a two-qubit
Hilbert space.}

Under the six-parameter local unitaries $U_A, U_B \in SU(2)$, these
coefficient matrices undergo the change
$C^{(k)} \rightarrow \widetilde{C}^{(k)} = U_A C^{(k)} U_B^T$, $k=1,2,3,4$.
We begin by using this local freedom to first bring $C^{(1)}$ to the
canonical form
\begin{equation}
\widetilde{C}^{(1)} = 
\begin{pmatrix}
    \cos \theta_1 & 0 \\
    0             & \sin \theta_1
\end{pmatrix},
\qquad 0 \le \theta_1 \le \pi/4,
\end{equation}
$\cos \theta_1$, $\sin \theta_1$ being respectively the larger and smaller singular
values of $C^{(1)}$.
In this process we have already used up all local unitary freedom 
{\em except conjugation
by diagonal $SU(2)$ matrices}: $U_A = {\rm diag}(e^{-i \eta}, e^{i \eta})$,
$U_B = U_A^{\star}$ [Just as we are free to multiply every
$|\Psi_k\rangle$ of a frame by a phase factor $e^{i\eta_k}$, so also
we can multiply every coefficient matrix by a unimodular scalar $e^{i\eta_k}$].

To obtain the canonical form for the second vector, note that any normalized matrix orthogonal to $\widetilde{C}^{(1)}$ is
necessarily of the form
\begin{equation}
\begin{pmatrix}
    \alpha \sin \theta_1 &
        \sqrt{1 - \alpha^2} e^{i\phi} \sin \theta_2 \\
    \sqrt{1 - \alpha^2} e^{i\phi'} \cos \theta_2 &
        -\alpha \cos \theta_1
\end{pmatrix},
\qquad 0 \le \alpha \le 1.
\end{equation}
Now we may use up the sixth and last local freedom to render the
phases of the off-diagonal elements equal.
Thus the canonical form for the second matrix is
\begin{equation}
\widetilde{C}^{(2)} = 
\begin{pmatrix}
    \alpha \sin \theta_1 &
        \sqrt{1 - \alpha^2} e^{i\phi} \sin \theta_2 \\
    \sqrt{1 - \alpha^2} e^{i\phi} \cos \theta_2 &
        -\alpha \cos \theta_1
\end{pmatrix}.
\end{equation}
With the local unitary freedom having been thus fully exhausted,
$\widetilde{C}^{(3)}$ has the canonical form
\begin{equation}
\widetilde{C}^{(3)} = 
\begin{pmatrix}
    \beta \sin \theta_1 &
        \sqrt{1 - \beta^2} e^{i\phi_3} \cos \theta_3 \\
    -\sqrt{1 - \beta^2} e^{i\phi_3'} \sin \theta_3 &
        -\beta \cos \theta_1
\end{pmatrix}.
\end{equation}
It should be noted that the four (real) parameters
$\beta, \theta_3, \phi_3, \phi_3'$
of $\widetilde{C}^{(3)}$ are not arbitrary.
While $\widetilde{C}^{(3)}$ is manifestly orthogonal to
$\widetilde{C}^{(1)}$, the orthogonality requirement
${\rm Tr}( \widetilde{C}^{(3)\dagger} \widetilde{C}^{(2)} ) = 0$
when enforced would determine these four parameters in terms of two
independent parameters.
Finally, $\widetilde{C}^{(4)}$ has the canonical form
\begin{equation}
\widetilde{C}^{(4)} = 
\begin{pmatrix}
    \gamma \sin \theta_1 &
        \sqrt{1 - \gamma^2} e^{i\phi_4} \cos \theta_4 \\
    -\sqrt{1 - \gamma^2} e^{i\phi_4'} \sin \theta_4 &
        -\gamma \cos \theta_1
\end{pmatrix},
\end{equation}
but it is clear that none of
$\gamma, \theta_4, \phi_4, \phi_4'$ is a free (continuous) parameter: they get
fixed by the two complex-valued conditions
$ {\rm Tr}( \widetilde{C}^{(4)\dagger} \widetilde{C}^{(2)} ) = 0
= {\rm Tr}( \widetilde{C}^{(4)\dagger} \widetilde{C}^{(3)} )$.

Returning to $\widetilde{C}^{(3)}$, the complex-valued condition 
${\rm Tr}( \widetilde{C}^{(3)\dagger} \widetilde{C}^{(2)} ) = 0$,
when written out in detail, reads
\begin{equation}
\begin{aligned}
\alpha \beta + \sqrt{ (1-\alpha^2) (1-\beta^2) }
& \left[ e^{i(\phi-\phi_3)} \sin \theta_2 \cos \theta_3 \right. \\
    &\left. - e^{i(\phi-\phi'_3)} \cos \theta_2 \sin \theta_3 \right] = 0.
\end{aligned}
\end{equation}
The imaginary part of this equation leads to the restriction
\begin{equation} \label{phi_prime_eqn}
\phi'_3 = \phi - \sin^{-1} \left[ \frac{\tan \theta_2} {\tan \theta_3}
    \sin (\phi - \phi_3) \right],
\end{equation}
while the real part requires
\begin{equation} \label{beta_eqn}
\beta = \sqrt{ \frac{ (1-\alpha^2) \Gamma^2 }
               { \alpha^2 + (1-\alpha^2) \Gamma^2 } },
\end{equation}
where $\Gamma = \sin \theta_2 \cos \theta_3 \cos(\phi-\phi_3)
- \cos \theta_2 \sin \theta_3 \cos(\phi-\phi'_3)$.
Thus in the present scheme we may choose the following six as free parameters:
$0 \le \alpha \le 1$, $0 \le \theta_1 \le \pi/4$,
$0 \le \theta_2, \theta_3 \le \pi/2$ and $0 \le \phi,\phi_3 < 2\pi$.
In terms of these six parameters the other two parameters for
$|\psi_3 \rangle$ or $\widetilde{C}^{(3)}$, namely $\phi_3', \beta$, can be determined through
equations (\ref{phi_prime_eqn}), (\ref{beta_eqn}).
Note that the allowed ranges for angles are not completely free and have to satisfy
constraints such that the argument of $\sin^{-1}$ in
Eqn.~(\ref{phi_prime_eqn}) has magnitude less than or equal to $1$, and
$\Gamma \le 0$ since $\alpha, \beta \ge 0$ by assumption.

Let us quickly do a parameter counting to check the reasonableness of
this parameterization.
A generic orthonormal frame in the two-qubit Hilbert space would be
expected to be
parameterized by twelve parameters: six (real, continuous) parameters for the first vector
(a generic element of $CP^3$), four for the second (an element of the orthogonal $CP^2$), two for the
third (the $CP^1 \sim S^2$ orthogonal to the first two vectors), and none for the fourth.
We have thus `efficiently' used the $3+3=6$-parameter local unitary
freedom to maximal effect to go
from twelve to six: $|\psi_1 \rangle$ is left with one parameter
($\theta_1$) with five local unitary parameters used up,
$|\psi_2 \rangle$ has three parameters ($\alpha, \theta_2, \phi$) with
the sixth and last local parameter used up,
$\langle \psi_1 | \psi_3 \rangle = \langle \psi_2 | \psi_3 \rangle = 0$
implies just two residual (continuous) parameters for $|\psi_3 \rangle$, namely ($\theta_3, \phi_3$).
And $|\psi_4 \rangle$ is automatically fixed by the requirement that
this  four-dimensional vector is orthogonal to
$|\psi_1 \rangle, |\psi_2 \rangle, |\psi_3 \rangle$.

Note that the special two-parameter family of frames or tetrahedra
discussed earlier corresponds to the choice $\alpha=1$, which
immediately renders $\beta = 0 = \gamma$.
Unlike the case of this special two-parameter family, the condition for
separability in the general case of six canonical parameters
does not break into direct sum of a pair of
$2 \times 2$ matrices.
And hence the resulting separable subsets of the associated tetrahedra can have
boundaries considerably more complex than quadratic and planar
surfaces of the earlier two-parameter case: they can be upto quadric surfaces.

For each $\xi \in \Gamma_{22}$ we have numerically evaluated the
fractional volume $f(\xi)$ of the convex subset of separable states in
$\Delta_3(\xi)$, and using this result in Eq.\,(1) we find:
\begin{itemize}
\item
$f(\xi) \ge 0.5$ for every $\xi \in \Gamma_{22}$, the inequality
saturating only for the Bell or magic frame (modulo local unitaries)
\item
the integral in Eq.~(\ref{vsep}) for $v^{\rm sep}$ actually evaluates to
the value 0.632, consistent with the earlier result of\,Ref.~\cite{Zyczkowski1998}.
\end{itemize}

\section{Monte Carlo Sampling: Higher Dimensional Systems}
\begin{figure}
  \centering
  \subfloat[$2 \times 2$ system]{\includegraphics[clip,width=0.92\linewidth]{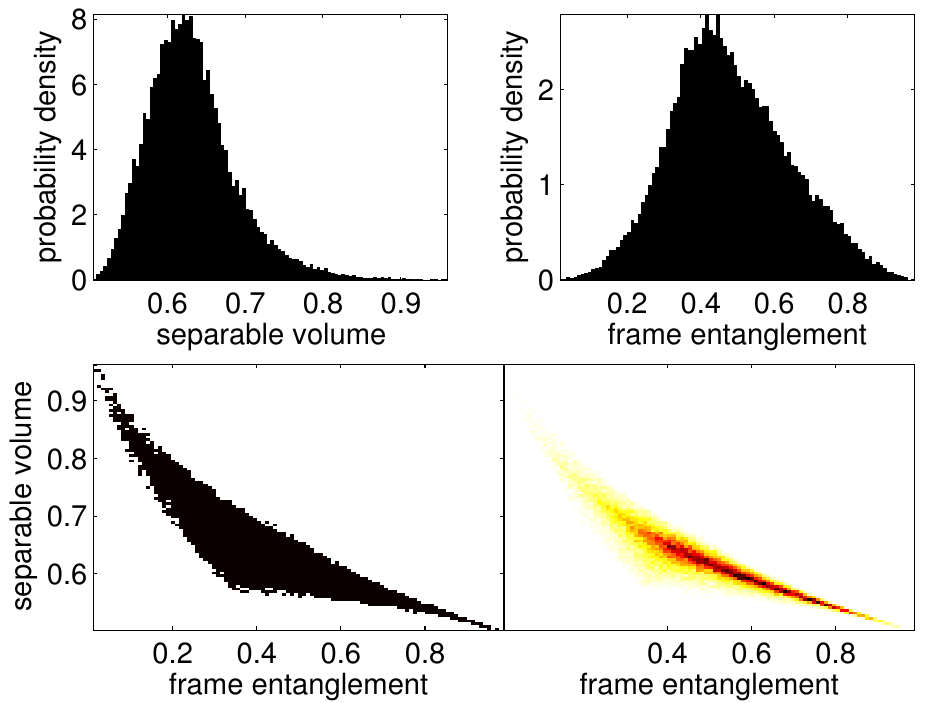}}
  \\
  \subfloat[$2 \times 3$ system]{\includegraphics[clip,width=0.92\linewidth]{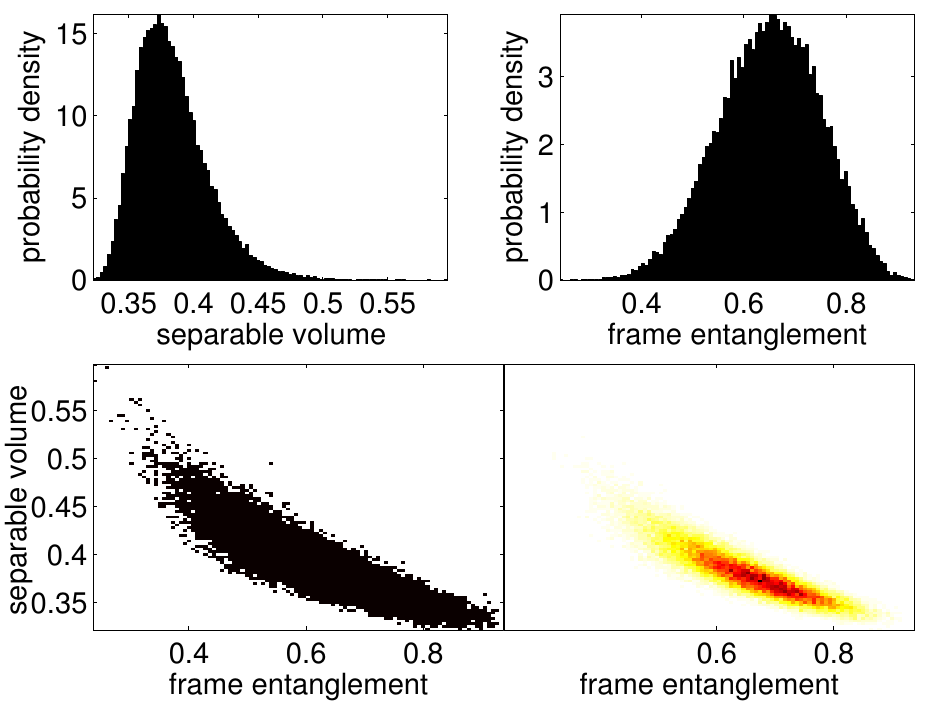}}
  \\
  \subfloat[$3 \times 3$ system]{\includegraphics[clip,width=0.92\linewidth]{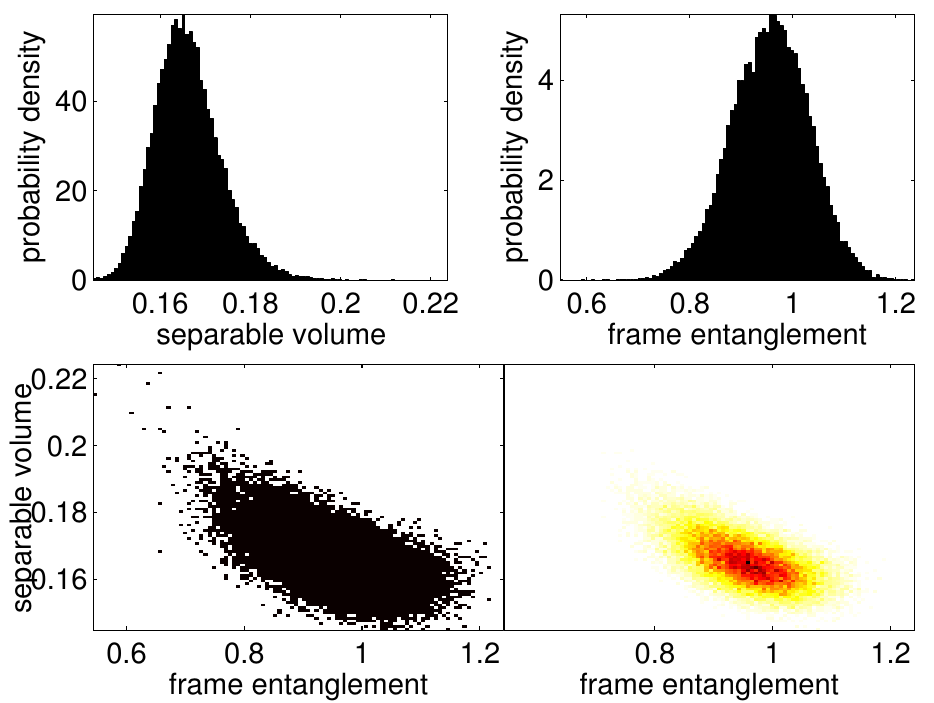}}
  \caption{(Color online) Distribution of separable volume and frame entanglement over the orbit of frames.
  In each subfigure the top-left plot shows the distribution of (fractional)
  separable volume, the top-right plot shows the distribution of
  frame entanglement, the bottom-left plot shows the scatter plot of
  all pairs of the separable volume and frame entanglement, and the bottom-right
  plot shows the 2-d histogram corresponding to the joint distribution
  of separable volume and frame entanglement.}
  \label{SeparableVolumeDtbn}
\end{figure}

\begin{figure}
  \centering
  \includegraphics[width=1.0\linewidth]{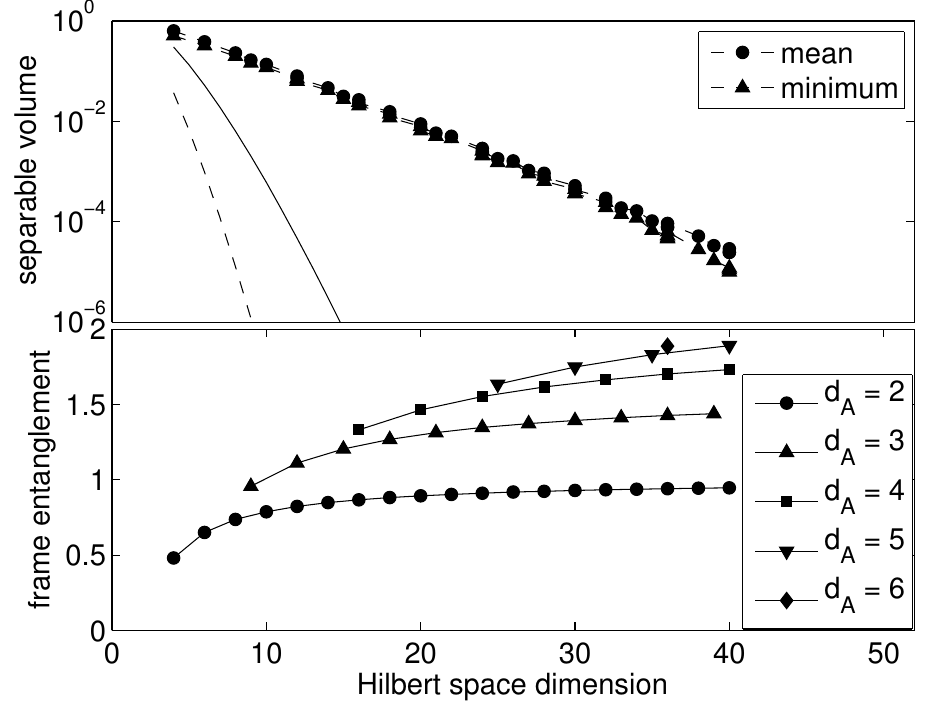}
  \caption{Top panel shows mean and minimum of separable volume over
  frames as a function of Hilbert space dimension, showing an
  exponential decrease.
  It also shows the lower bounds given by~\cite{Zyczkowski1998} as
  solid line, and the one  given by~\cite{Vidal1999} as dashed line.
  Bottom panel shows the corresponding mean frame entanglement with
  different symbols for different $d_A$ for fixed $d_A d_B$.}
  \label{MeanVsSystemSize}
\end{figure}

\begin{table}
\caption{\label{tab:volume} Relative separable volume 
 for bipartite  systems with fixed total
Hilbert space dimensions $d_Ad_B$ that can be decomposed 
as ${\cal H}_A\otimes {\cal H}_B$ in more ways than one.}
\begin{tabular}{|c|l|c|c|}
\hline
$d_A d_B$ & $d_A \times d_B$ & \quad Mean \quad & Minimum \\
\hline
$12$ & $2 \times  6$ & $0.0796$ & $0.0708$ \\
     & $3 \times  4$ & $0.0724$ & $0.0631$ \\
\hline
$16$ & $2 \times  8$ & $0.0268$ & $0.0242$ \\
     & $4 \times  4$ & $0.0233$ & $0.0204$ \\
\hline
$18$ & $2 \times  9$ & $0.0154$ & $0.0140$ \\
     & $3 \times  6$ & $0.0135$ & $0.0118$ \\
\hline
$20$ & $2 \times 10$ & $0.0088$ & $0.0080$ \\
     & $4 \times  5$ & $0.0075$ & $0.0065$ \\
\hline
$24$ & $2 \times 12$ & $0.0029$ & $0.0026$ \\
     & $3 \times  8$ & $0.0025$ & $0.0021$ \\
     & $4 \times  6$ & $0.0024$ & $0.0021$ \\
\hline
\end{tabular}
\end{table}

To gain quick insight into the situation in respect of higher
dimensional systems we perform Monte Carlo sampling of the sets $\Gamma_{AB}$ and
$\Delta_{{\,d}_Ad_B-1}$ following the scheme in~\cite{Zyczkowski1998}.
However, instead of sampling from the joint distribution we estimate
the relative separable volume for each frame. The relative separable volume in the full space is simply the average
over frames, as expressed in Eq.~(\ref{vsep}).
For most systems, $2^{15} \approx 3 \times 10^4$ frames were sampled
from $\Gamma_{AB}$ using Haar measure, and for each frame $10^6$ points
were sampled from the corresponding simplex $\Delta_{{\,d}_Ad_B-1}$ uniformly.
Although the Haar measure for the orbit $\Gamma_{AB}$ is the
natural one, there is no `unique' measure to sample the simplex
$\Delta_{{\,d}_Ad_B-1}$, and indeed different measures have been
motivated and used in\,\cite{Zyczkowski1999,Slater1999}. However, we have
used the uniform measure to be consistent with the work of 
\.{Z}yczkowski {\em et. al.}\,\cite{Zyczkowski1998} which motivated the 
present work. 
Fig.~\ref{SeparableVolumeDtbn} shows the distribution of relative
separable volume and {\em frame entanglement}, the average
entanglement of the orthonormal pure states ($d_A d_B$ in number)
defining the frame.
It also shows the joint distribution of these two quantities as well
as their scatter plots.
We observe that for $2 \times 2$ system the separable volume
distribution becomes narrow as the frame entanglement approaches $1$,
which does not happen for other cases.
This is possibly a consequence of the fact that for a $2 \times 2$ system
there exists only one maximally entangled frame modulo local unitary,
whereas for higher dimensional systems there are many locally
inequivalent maximally entangled frames~\cite{Vollbrecht2000,Werner2001}.
We show in Fig.~\ref{MeanVsSystemSize} the mean and minimum separable volume and frame
entanglement as a function of Hilbert space dimension.
Consistent with earlier work~\cite{Zyczkowski1998}, we find that the
separable volume decreases exponentially with Hilbert space dimension.
Systems with the same total or composite Hilbert space dimension but
different subsystem dimensions have only slightly different
separable volume which is not prominently visible in
Fig.~\ref{MeanVsSystemSize}, and so has been detailed in
Table~\ref{tab:volume}.

Our approach generalizes to higher dimensional systems, wherein
qualitatively new features emerge.
For instance, for the qutrit-qutrit systems not all frames of
maximally entangled states are local unitarily  equivalent and, consequently,
they lead to unequal fractional volume of separable states and,
perhaps surprisingly, the `Bell frame' is not the one to result in minimum
separable volume.
This result is significant should it possibly imply that for higher
dimensional $d \times d$ systems, the Bell frame is not the most
robust one among the maximally entangled frames.

To indicate what we mean by Bell frame for a $d \times d$ system,
define a pair of $d \times d$ matrices $X, Y$ through
$X = {\rm diag}\,(1, \omega_d, \omega_d^2, \cdots, \omega_d^{d-1})$,
$Y_{jk} = \delta_{j+1,k}$ where $\omega_d = \exp(-i 2 \pi /d)$ and
$j+1 = k$ is to be understood in the $mod\ d$ sense.
It is clear that the $d^2$ matrices
$C^{\alpha \beta} = d^{-1/2} X^{\alpha} Y^{\beta}$,
$\alpha, \beta = 1, 2, \ldots, d$ viewed as coefficient matrices in
the computational product basis correspond to maximally entangled
orthonormal vectors.
For brevity we call this basis of maximally entangled states `the Bell frame'.

\section{Qubit-Qutrit System}
Analogous to the special two-parameter family of the two-qubit frames
considered earlier, we now consider a special three-parameter family
of orthogonal frames for the $2 \times 3$ system representing a
qubit-qutrit system:
\begin{equation}
\begin{aligned}
& C_{|\psi_1 \rangle} = 
\begin{pmatrix}
    \cos \theta & 0           & 0 \\
              0 & \sin \theta & 0 
\end{pmatrix},
C_{|\psi_2 \rangle} = 
\begin{pmatrix}
    \sin \theta & 0            & 0 \\
              0 & -\cos \theta & 0
\end{pmatrix}, \\
& C_{|\psi_3 \rangle} = 
\begin{pmatrix}
    0 &  \cos \alpha & 0          \\
    0 &            0 & \sin \alpha
\end{pmatrix},
C_{|\psi_4 \rangle} = 
\begin{pmatrix}
   0 &  \sin \alpha & 0           \\
   0 &            0 & -\cos \alpha
\end{pmatrix}, \\
& C_{|\psi_5 \rangle} = 
\begin{pmatrix}
             0 & 0 & \cos \beta \\
    \sin \beta & 0 & 0          
\end{pmatrix},
C_{|\psi_6 \rangle} = 
\begin{pmatrix}
             0 & 0 & \sin \beta \\
   -\cos \beta & 0 & 0          
\end{pmatrix}.
\end{aligned}
\end{equation}
The particular case $\theta = \alpha = \beta = \pi/4$ may be called
the Bell basis for the $2 \times 3$ system.
We find using Monte Carlo sampling, that the relative
separable volume is approximately $0.377$.
We find numerically that there are other maximally entangled frames
which do not belong to this special parameterization {\em that have
lower separable volume than the Bell-diagonal frame}.

\section{Volume of Hyperspheres}
\begin{figure}
\centering
\includegraphics[width=1.0\linewidth]{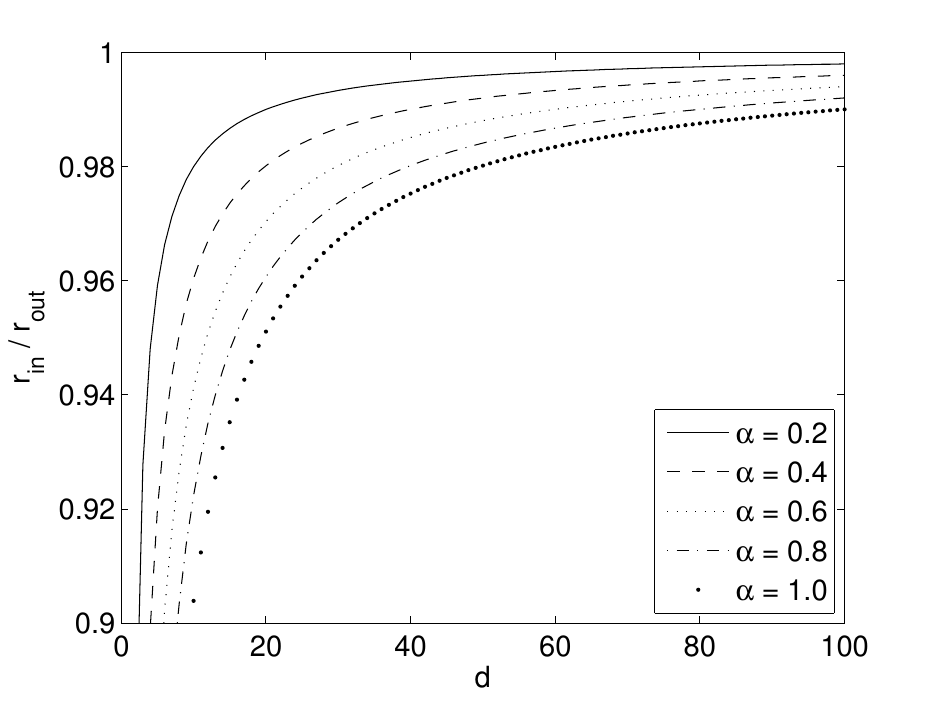}
\caption{Behaviour of the ratio of radii for hyperspheres in 
state space $\Omega_d$ with Hilbert space
dimension $d$, assuming exponential decrease $e^{-\alpha d}$ in the
ratio of volumes.}
\label{hypersphere}
\end{figure}
In this Section we suggest that an exponential decrease in the volume of the
separable states with increasing Hilbert space dimension implies an
{\em increase} in `effective radius' for separable
states~\cite{Zyczkowski1998,Vidal1999,Szarek2005}.
To gain some perspective, let us discuss the effect of Hilbert space 
dimension on the ratio of volumes of hyperspheres (in the state space) with constant
ratio of radii.
Let $r_{\rm sep}$ and $r_{\rm tot}$ be the radii of the inner
(separable) and outer (total)
hyperspheres respectively ($r_{\rm sep} < r_{\rm tot}$).
The ratio of the volumes of these hyperspheres in $n$ dimensions is
$V^{\rm sep} / V^{\rm tot} = (r_{\rm sep} / r_{\rm tot})^n$.
Thus the ratio of the volumes decreases exponentially with the
dimension $n$ even if the ratio of the radii is constant.
For a quantum system represented by a $d$-dimensional Hilbert space,
the state space, i.e. the space of density matrices, is $d^2-1$
dimensional.
Thus, in the hypothetical case in which the set of all states and the
set of separable states were (concentric) hyperspheres with ratio of 
radii independent of $d$, the ratio of the volume would have decreased as
$(r_{\rm sep} / r_{\rm tot})^{(d^2-1)}$, which is faster than exponential in 
Hilbert space dimension $d$.
Thus, if one observes no-stronger-than exponential decrease ($e^{-\alpha d}$)
in the actual ratio of volumes, then the ratio
$(r_{\rm sep} / r_{\rm tot})$ ought to 
increase with $d$ as
$e^{-\alpha d/(d^2-1)}$, and approach $1$ asymptotically (see Fig.~\ref{hypersphere}).
Thus it appears that a no-faster-than exponential decrease in relative separable
volume with Hilbert space dimension implies that the ``effective''
relative radius of the separable region must actually {\em increase} with
dimension.
This seems to be a new insight, as earlier results have claimed a
decreasing lower bound on this effective radius~\cite{Szarek2005}.
More importantly, there exists one claim that {\em an upper
bound} on this effective radius too decreases with increasing Hilbert
space dimension~\cite{Szarek2005} for the case of quantum systems
composed of many qubits.

\section{Summary and Conclusions}
In this paper we have analyzed in some detail the geometry of
separable states in some three-sections of the $15$-parameter
two-qubit state space $\Omega_4$, and some of these sections were
pictured in Fig.~\ref{SeparableSet}.
This hopefully gives some insight into the geometry of separable sets for
two-qubits.
We have also given a general parameterization for the state
space of two-qubit system.
We believe our analysis shows why the surprising result of
Ref.~\cite{Zyczkowski1998} could indeed have been `anticipated'.
Using Monte Carlo sampling of the state space of the higher
dimensional system, we have explored the relation between separable
volume and frame entanglement.
One of the major surprising results is that for higher dimensional
systems Bell frame is not the one having minimum separable
volume.
This result could possibly have important consequences for generating robust
entangled states.
We have also pointed out that a no-stronger-than
 exponential decrease in relative separable volume
with Hilbert space dimension actually implies an increase in the
`effective radius' of the separable set, contrary to earlier claims.

Though we have considered the uniform measure on the simplex, other
measures can also be considered.
As an example we find that with Dirichlet measure ($\nu = 1/2$) the
separable volumes are 0.350 ($2 \times 2$), 0.122 ($2 \times 3$) and
0.022 ($3 \times 3$) consistent with earlier
results~\cite{Slater1999,Zyczkowski1999}.
The computational cost of our approach appears to grow as
$\sim d^{5/2}$ with Hilbert space dimension $d (=d_A d_B)$, which
makes it possible to go to even bigger systems if sufficient
computational resources are available.
Since the Monte Carlo method employed is embarrassingly parallel the
performance of simulation should increase linearly with number of
available processors.

\vskip 0.3cm

\noindent
{\bf Note}: Part of this work was presented in the Asian Quantum Information 
Science Conference AQIS'13. 
Computations were carried out at the ``Annapurna'' supercomputer 
facility of The Institute of Mathematical Sciences.
}

\begin{thebibliography}{20}%
\makeatletter
\providecommand \@ifxundefined [1]{%
 \@ifx{#1\undefined}
}%
\providecommand \@ifnum [1]{%
 \ifnum #1\expandafter \@firstoftwo
 \else \expandafter \@secondoftwo
 \fi
}%
\providecommand \@ifx [1]{%
 \ifx #1\expandafter \@firstoftwo
 \else \expandafter \@secondoftwo
 \fi
}%
\providecommand \natexlab [1]{#1}%
\providecommand \enquote  [1]{``#1''}%
\providecommand \bibnamefont  [1]{#1}%
\providecommand \bibfnamefont [1]{#1}%
\providecommand \citenamefont [1]{#1}%
\providecommand \href@noop [0]{\@secondoftwo}%
\providecommand \href [0]{\begingroup \@sanitize@url \@href}%
\providecommand \@href[1]{\@@startlink{#1}\@@href}%
\providecommand \@@href[1]{\endgroup#1\@@endlink}%
\providecommand \@sanitize@url [0]{\catcode `\\12\catcode `\$12\catcode
  `\&12\catcode `\#12\catcode `\^12\catcode `\_12\catcode `\%12\relax}%
\providecommand \@@startlink[1]{}%
\providecommand \@@endlink[0]{}%
\providecommand \url  [0]{\begingroup\@sanitize@url \@url }%
\providecommand \@url [1]{\endgroup\@href {#1}{\urlprefix }}%
\providecommand \urlprefix  [0]{URL }%
\providecommand \Eprint [0]{\href }%
\providecommand \doibase [0]{http://dx.doi.org/}%
\providecommand \selectlanguage [0]{\@gobble}%
\providecommand \bibinfo  [0]{\@secondoftwo}%
\providecommand \bibfield  [0]{\@secondoftwo}%
\providecommand \translation [1]{[#1]}%
\providecommand \BibitemOpen [0]{}%
\providecommand \bibitemStop [0]{}%
\providecommand \bibitemNoStop [0]{.\EOS\space}%
\providecommand \EOS [0]{\spacefactor3000\relax}%
\providecommand \BibitemShut  [1]{\csname bibitem#1\endcsname}%
\let\auto@bib@innerbib\@empty
\bibitem [{\citenamefont {Bengtsson}\ and\ \citenamefont
  {\.{Z}yczkowski}(2006)}]{Bengtsson2006}%
  \BibitemOpen
  \bibfield  {author} {\bibinfo {author} {\bibfnamefont {I.}~\bibnamefont
  {Bengtsson}}\ and\ \bibinfo {author} {\bibfnamefont {K.}~\bibnamefont
  {\.{Z}yczkowski}},\ }\href {\doibase 10.1017/CBO9780511535048} {\emph
  {\bibinfo {title} {{Geometry of Quantum States}}}}\ (\bibinfo  {publisher}
  {Cambridge University Press},\ \bibinfo {address} {Cambridge},\ \bibinfo
  {year} {2006})\BibitemShut {NoStop}%
\bibitem [{\citenamefont {Bengtsson}\ \emph {et~al.}(2013)\citenamefont
  {Bengtsson}, \citenamefont {Weis},\ and\ \citenamefont
  {\.{Z}yczkowski}}]{Bengtsson2013}%
  \BibitemOpen
  \bibfield  {author} {\bibinfo {author} {\bibfnamefont {I.}~\bibnamefont
  {Bengtsson}}, \bibinfo {author} {\bibfnamefont {S.}~\bibnamefont {Weis}}, \
  and\ \bibinfo {author} {\bibfnamefont {K.}~\bibnamefont {\.{Z}yczkowski}},\
  }in\ \href {\doibase 10.1007/978-3-0348-0448-6\_15} {\emph {\bibinfo
  {booktitle} {Geometric Methods in Physics}}},\ \bibinfo {editor} {edited by\
  \bibinfo {editor} {\bibfnamefont {P.}~\bibnamefont {Kielanowski}}, \bibinfo
  {editor} {\bibfnamefont {S.~T.}\ \bibnamefont {Ali}}, \bibinfo {editor}
  {\bibfnamefont {A.}~\bibnamefont {Odzijewicz}}, \bibinfo {editor}
  {\bibfnamefont {M.}~\bibnamefont {Schlichenmaier}}, \ and\ \bibinfo {editor}
  {\bibfnamefont {T.}~\bibnamefont {Voronov}}}\ (\bibinfo  {publisher}
  {Springer Basel},\ \bibinfo {year} {2013})\ pp.\ \bibinfo {pages}
  {175--197}\BibitemShut {NoStop}%
\bibitem [{\citenamefont {Goyal}\ \emph {et~al.}(2011)\citenamefont {Goyal},
  \citenamefont {Simon}, \citenamefont {Singh},\ and\ \citenamefont
  {Simon}}]{Goyal2011}%
  \BibitemOpen
  \bibfield  {author} {\bibinfo {author} {\bibfnamefont {S.~K.}\ \bibnamefont
  {Goyal}}, \bibinfo {author} {\bibfnamefont {B.~N.}\ \bibnamefont {Simon}},
  \bibinfo {author} {\bibfnamefont {R.}~\bibnamefont {Singh}}, \ and\ \bibinfo
  {author} {\bibfnamefont {S.}~\bibnamefont {Simon}},\ }\href@noop {} {}
  (\bibinfo {year} {2011}),\ \Eprint {http://arxiv.org/abs/1111.4427}
  {arXiv:1111.4427} \BibitemShut {NoStop}%
\bibitem [{\citenamefont {Sarbicki}\ and\ \citenamefont
  {Bengtsson}(2013)}]{Sarbicki2013}%
  \BibitemOpen
  \bibfield  {author} {\bibinfo {author} {\bibfnamefont {G.}~\bibnamefont
  {Sarbicki}}\ and\ \bibinfo {author} {\bibfnamefont {I.}~\bibnamefont
  {Bengtsson}},\ }\href {\doibase 10.1088/1751-8113/46/3/035306} {\bibfield
  {journal} {\bibinfo  {journal} {J. Phys. A}\ }\textbf {\bibinfo {volume}
  {46}},\ \bibinfo {pages} {035306} (\bibinfo {year} {2013})}\BibitemShut
  {NoStop}%
\bibitem [{\citenamefont {Horodecki}\ \emph {et~al.}(2009)\citenamefont
  {Horodecki}, \citenamefont {Horodecki},\ and\ \citenamefont
  {Horodecki}}]{Horodecki2009}%
  \BibitemOpen
  \bibfield  {author} {\bibinfo {author} {\bibfnamefont {R.}~\bibnamefont
  {Horodecki}}, \bibinfo {author} {\bibfnamefont {M.}~\bibnamefont
  {Horodecki}}, \ and\ \bibinfo {author} {\bibfnamefont {K.}~\bibnamefont
  {Horodecki}},\ }\href {\doibase 10.1103/RevModPhys.81.865} {\bibfield
  {journal} {\bibinfo  {journal} {Rev. Mod. Phys.}\ }\textbf {\bibinfo {volume}
  {81}},\ \bibinfo {pages} {865} (\bibinfo {year} {2009})}\BibitemShut
  {NoStop}%
\bibitem [{\citenamefont {\.{Z}yczkowski}\ \emph {et~al.}(1998)\citenamefont
  {\.{Z}yczkowski}, \citenamefont {Horodecki}, \citenamefont {Sanpera},\ and\
  \citenamefont {Lewenstein}}]{Zyczkowski1998}%
  \BibitemOpen
  \bibfield  {author} {\bibinfo {author} {\bibfnamefont {K.}~\bibnamefont
  {\.{Z}yczkowski}}, \bibinfo {author} {\bibfnamefont {P.}~\bibnamefont
  {Horodecki}}, \bibinfo {author} {\bibfnamefont {A.}~\bibnamefont {Sanpera}},
  \ and\ \bibinfo {author} {\bibfnamefont {M.}~\bibnamefont {Lewenstein}},\
  }\href {\doibase 10.1103/PhysRevA.58.883} {\bibfield  {journal} {\bibinfo
  {journal} {Phys. Rev. A}\ }\textbf {\bibinfo {volume} {58}},\ \bibinfo
  {pages} {883} (\bibinfo {year} {1998})}\BibitemShut {NoStop}%
\bibitem [{\citenamefont {\.{Z}yczkowski}(1999)}]{Zyczkowski1999}%
  \BibitemOpen
  \bibfield  {author} {\bibinfo {author} {\bibfnamefont {K.}~\bibnamefont
  {\.{Z}yczkowski}},\ }\href {\doibase 10.1103/PhysRevA.60.3496} {\bibfield
  {journal} {\bibinfo  {journal} {Phys. Rev. A}\ }\textbf {\bibinfo {volume}
  {60}},\ \bibinfo {pages} {3496} (\bibinfo {year} {1999})}\BibitemShut
  {NoStop}%
\bibitem [{\citenamefont {Slater}(1999)}]{Slater1999}%
  \BibitemOpen
  \bibfield  {author} {\bibinfo {author} {\bibfnamefont {P.~B.}\ \bibnamefont
  {Slater}},\ }\href {\doibase 10.1088/0305-4470/32/28/306} {\bibfield
  {journal} {\bibinfo  {journal} {J. Phys. A}\ }\textbf {\bibinfo {volume}
  {32}},\ \bibinfo {pages} {5261} (\bibinfo {year} {1999})}\BibitemShut
  {NoStop}%
\bibitem [{\citenamefont {Kendon}\ \emph
  {et~al.}(2002{\natexlab{a}})\citenamefont {Kendon}, \citenamefont
  {\.{Z}yczkowski},\ and\ \citenamefont {Munro}}]{Kendon2002a}%
  \BibitemOpen
  \bibfield  {author} {\bibinfo {author} {\bibfnamefont {V.}~\bibnamefont
  {Kendon}}, \bibinfo {author} {\bibfnamefont {K.}~\bibnamefont
  {\.{Z}yczkowski}}, \ and\ \bibinfo {author} {\bibfnamefont {W.}~\bibnamefont
  {Munro}},\ }\href {\doibase 10.1103/PhysRevA.66.062310} {\bibfield  {journal}
  {\bibinfo  {journal} {Phys. Rev. A}\ }\textbf {\bibinfo {volume} {66}},\
  \bibinfo {pages} {062310} (\bibinfo {year} {2002}{\natexlab{a}})}\BibitemShut
  {NoStop}%
\bibitem [{\citenamefont {Vidal}\ and\ \citenamefont
  {Tarrach}(1999)}]{Vidal1999}%
  \BibitemOpen
  \bibfield  {author} {\bibinfo {author} {\bibfnamefont {G.}~\bibnamefont
  {Vidal}}\ and\ \bibinfo {author} {\bibfnamefont {R.}~\bibnamefont
  {Tarrach}},\ }\href {\doibase 10.1103/PhysRevA.59.141} {\bibfield  {journal}
  {\bibinfo  {journal} {Phys. Rev. A}\ }\textbf {\bibinfo {volume} {59}},\
  \bibinfo {pages} {141} (\bibinfo {year} {1999})}\BibitemShut {NoStop}%
\bibitem [{\citenamefont {Verstraete}\ \emph {et~al.}(2001)\citenamefont
  {Verstraete}, \citenamefont {Audenaert},\ and\ \citenamefont {{De
  Moor}}}]{Verstraete2001}%
  \BibitemOpen
  \bibfield  {author} {\bibinfo {author} {\bibfnamefont {F.}~\bibnamefont
  {Verstraete}}, \bibinfo {author} {\bibfnamefont {K.}~\bibnamefont
  {Audenaert}}, \ and\ \bibinfo {author} {\bibfnamefont {B.}~\bibnamefont {{De
  Moor}}},\ }\href {\doibase 10.1103/PhysRevA.64.012316} {\bibfield  {journal}
  {\bibinfo  {journal} {Phys. Rev. A}\ }\textbf {\bibinfo {volume} {64}},\
  \bibinfo {pages} {012316} (\bibinfo {year} {2001})}\BibitemShut {NoStop}%
\bibitem [{\citenamefont {Slater}\ and\ \citenamefont
  {Dunkl}(2012)}]{Slater2012}%
  \BibitemOpen
  \bibfield  {author} {\bibinfo {author} {\bibfnamefont {P.~B.}\ \bibnamefont
  {Slater}}\ and\ \bibinfo {author} {\bibfnamefont {C.~F.}\ \bibnamefont
  {Dunkl}},\ }\href {\doibase 10.1088/1751-8113/45/9/095305} {\bibfield
  {journal} {\bibinfo  {journal} {J. Phys. A}\ }\textbf {\bibinfo {volume}
  {45}},\ \bibinfo {pages} {095305} (\bibinfo {year} {2012})}\BibitemShut
  {NoStop}%
\bibitem [{\citenamefont {Slater}(2013)}]{Slater2013}%
  \BibitemOpen
  \bibfield  {author} {\bibinfo {author} {\bibfnamefont {P.~B.}\ \bibnamefont
  {Slater}},\ }\href {\doibase 10.1088/1751-8113/46/44/445302} {\bibfield
  {journal} {\bibinfo  {journal} {J. Phys. A}\ }\textbf {\bibinfo {volume}
  {46}},\ \bibinfo {pages} {445302} (\bibinfo {year} {2013})},\ \Eprint
  {http://arxiv.org/abs/1301.6617} {1301.6617} \BibitemShut {NoStop}%
\bibitem [{\citenamefont {Kendon}\ \emph
  {et~al.}(2002{\natexlab{b}})\citenamefont {Kendon}, \citenamefont {Nemoto},\
  and\ \citenamefont {Munro}}]{Kendon2002}%
  \BibitemOpen
  \bibfield  {author} {\bibinfo {author} {\bibfnamefont {V.}~\bibnamefont
  {Kendon}}, \bibinfo {author} {\bibfnamefont {K.}~\bibnamefont {Nemoto}}, \
  and\ \bibinfo {author} {\bibfnamefont {W.}~\bibnamefont {Munro}},\ }\href
  {\doibase 10.1080/09500340110120914} {\bibfield  {journal} {\bibinfo
  {journal} {J. Mod. Opt.}\ }\textbf {\bibinfo {volume} {49}},\ \bibinfo
  {pages} {1709} (\bibinfo {year} {2002}{\natexlab{b}})}\BibitemShut {NoStop}%
\bibitem [{\citenamefont {Lang}\ and\ \citenamefont {Caves}(2010)}]{Lang2010}%
  \BibitemOpen
  \bibfield  {author} {\bibinfo {author} {\bibfnamefont {M.~D.}\ \bibnamefont
  {Lang}}\ and\ \bibinfo {author} {\bibfnamefont {C.~M.}\ \bibnamefont
  {Caves}},\ }\href {\doibase 10.1103/PhysRevLett.105.150501} {\bibfield
  {journal} {\bibinfo  {journal} {Phys. Rev. Lett.}\ }\textbf {\bibinfo
  {volume} {105}},\ \bibinfo {pages} {150501} (\bibinfo {year}
  {2010})}\BibitemShut {NoStop}%
\bibitem [{\citenamefont {Peres}(1996)}]{Peres1996}%
  \BibitemOpen
  \bibfield  {author} {\bibinfo {author} {\bibfnamefont {A.}~\bibnamefont
  {Peres}},\ }\href {\doibase 10.1103/PhysRevLett.77.1413} {\bibfield
  {journal} {\bibinfo  {journal} {Phys. Rev. Lett.}\ }\textbf {\bibinfo
  {volume} {77}},\ \bibinfo {pages} {1413} (\bibinfo {year}
  {1996})}\BibitemShut {NoStop}%
\bibitem [{\citenamefont {Horodecki}\ \emph {et~al.}(1996)\citenamefont
  {Horodecki}, \citenamefont {Horodecki},\ and\ \citenamefont
  {Horodecki}}]{Horodecki1996}%
  \BibitemOpen
  \bibfield  {author} {\bibinfo {author} {\bibfnamefont {M.}~\bibnamefont
  {Horodecki}}, \bibinfo {author} {\bibfnamefont {P.}~\bibnamefont
  {Horodecki}}, \ and\ \bibinfo {author} {\bibfnamefont {R.}~\bibnamefont
  {Horodecki}},\ }\href {\doibase 10.1016/S0375-9601(96)00706-2} {\bibfield
  {journal} {\bibinfo  {journal} {Physics Letters A}\ }\textbf {\bibinfo
  {volume} {223}},\ \bibinfo {pages} {1} (\bibinfo {year} {1996})}\BibitemShut
  {NoStop}%
\bibitem [{\citenamefont {Vollbrecht}\ and\ \citenamefont
  {Werner}(2000)}]{Vollbrecht2000}%
  \BibitemOpen
  \bibfield  {author} {\bibinfo {author} {\bibfnamefont {K.~G.~H.}\
  \bibnamefont {Vollbrecht}}\ and\ \bibinfo {author} {\bibfnamefont {R.~F.}\
  \bibnamefont {Werner}},\ }\href {\doibase 10.1063/1.1286032} {\bibfield
  {journal} {\bibinfo  {journal} {J. Math. Phys.}\ }\textbf {\bibinfo {volume}
  {41}},\ \bibinfo {pages} {6772} (\bibinfo {year} {2000})}\BibitemShut
  {NoStop}%
\bibitem [{\citenamefont {Werner}(2001)}]{Werner2001}%
  \BibitemOpen
  \bibfield  {author} {\bibinfo {author} {\bibfnamefont {R.~F.}\ \bibnamefont
  {Werner}},\ }\href {\doibase 10.1088/0305-4470/34/35/332} {\bibfield
  {journal} {\bibinfo  {journal} {J. Phys. A}\ }\textbf {\bibinfo {volume}
  {34}},\ \bibinfo {pages} {7081} (\bibinfo {year} {2001})}\BibitemShut
  {NoStop}%
\bibitem [{\citenamefont {Szarek}(2005)}]{Szarek2005}%
  \BibitemOpen
  \bibfield  {author} {\bibinfo {author} {\bibfnamefont {S.}~\bibnamefont
  {Szarek}},\ }\href {\doibase 10.1103/PhysRevA.72.032304} {\bibfield
  {journal} {\bibinfo  {journal} {Phys. Rev. A}\ }\textbf {\bibinfo {volume}
  {72}},\ \bibinfo {pages} {032304} (\bibinfo {year} {2005})}\BibitemShut
  {NoStop}%
\end{thebibliography}
%
\end{document}